\def \araa {{\rm {ARA\&A}}}
\def \apj {{\rm {ApJ}}}

\def \mnras {{\rm {MNRAS}}}

\def \solyr {{\rm M$_\odot$\,yr$^{-1}$}}

\def \micron{{\rm \,$\mu$m}}

\documentclass{iau} 
\usepackage{graphicx}

\title[Star formation rates on global and cloud scales within the Galactic Centre] 
{Star formation rates on global and cloud scales within the Galactic Centre}

\author[Barnes, Longmore, Battersby, Bally, Kruijssen]   
{A.T. Barnes$^{1}$, S.N. Longmore$^{1}$, C. Battersby$^{2}$, J. Bally$^{3}$, J.M.D. Kruijssen$^{4,5}$ \\
}

\affiliation{$^{1}$Astrophysics Research Institute, Liverpool John Moores University\\
 $^{2}$Harvard-Smithsonian Center for Astrophysics\\
 $^{3}$Centre for Astrophysics and Space Astronomy, University of Colorado \\
 $^{4}$Astronomisches Rechen-Institut, Zentrum f\"{u}r Astronomie der Universit\"{a}t Heidelberg \\ 
$^5$Max-Planck Institut f\"{u}r Astronomie\\
}

\pubyear{2016}
\volume{322}  
\jname{`The Multi-Messenger Astrophysics of the Galactic Centre'}
\editors{R. Crocker, S. Longmore, G. Bicknell}
\begin{document}

\maketitle

\vspace{-2mm}
\begin{abstract}
The environment within the inner few hundred parsecs of the Milky Way, known as the ``Central Molecular Zone'' (CMZ), harbours densities and pressures orders of magnitude higher than the Galactic Disc; akin to that at the peak of cosmic star formation \cite[(Kruijssen \& Longmore 2013)]{kruijssen_2013}. Previous studies have shown that current theoretical star-formation models under-predict the observed level of star-formation (SF) in the CMZ by an order of magnitude given the large reservoir of dense gas it contains. Here we explore potential reasons for this apparent dearth of star formation activity.

\end{abstract}

\firstsection 

\vspace{-6mm}
\section{Large scale: star formation rates on 100\,pc scales}

\begin{figure}[b]
\begin{center}
 \includegraphics[trim = 4mm 1mm 1mm 1mm, clip,angle=0,width=0.95\textwidth]{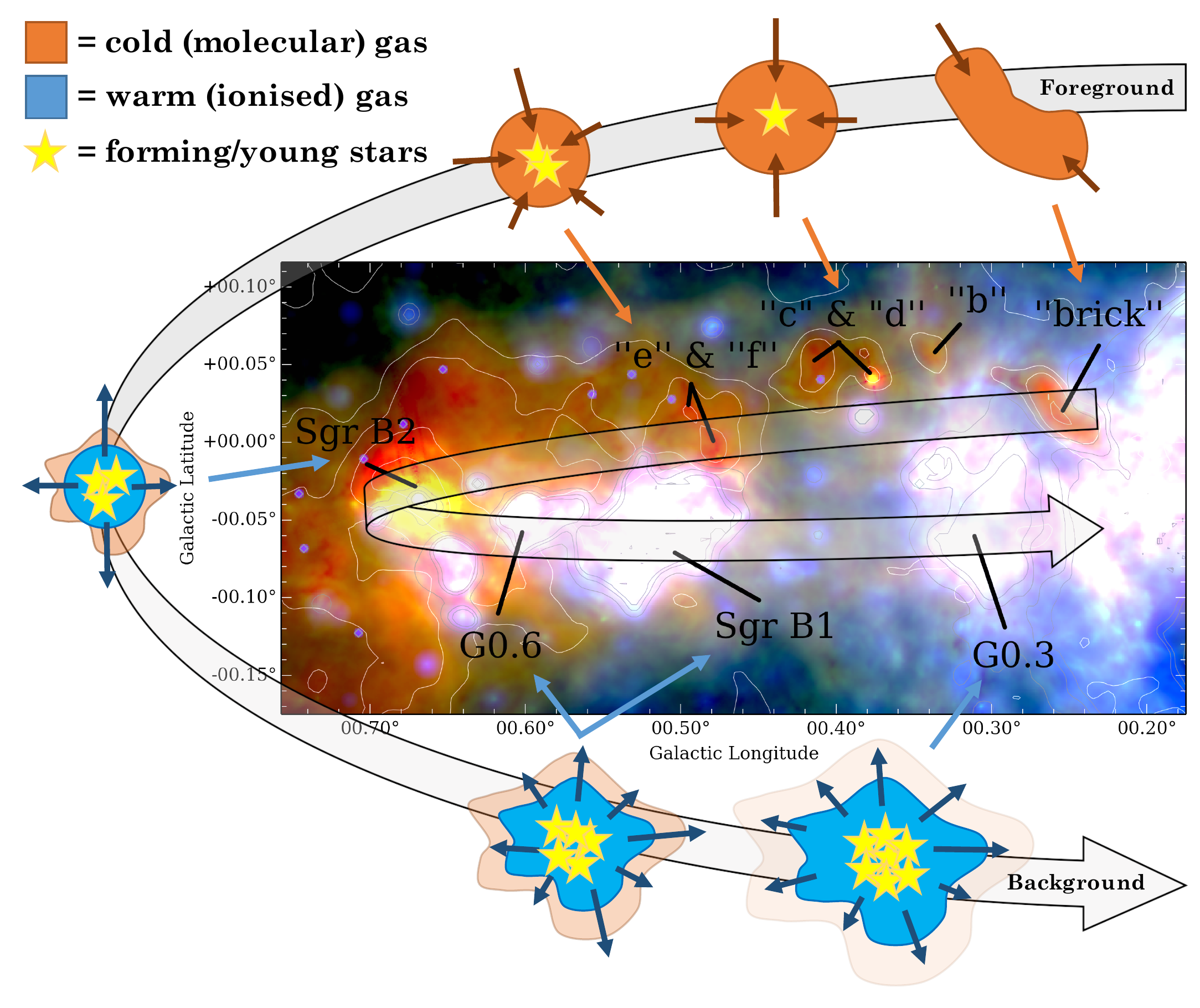} 
\vspace{-4mm}
 \caption{Three colour image of the sources considered in section\,\ref{small_scale}. Shown in blue and red are warm and cold gas, determined from a two component fit to spectral energy distribution between $5.8 - 500$\micron, and in green is the 70\,\micron\ emission. Overlaid are contours of the warm and cool gas column in grey and white, respectively.  Overlaid as is a top-down schematic diagram of the orbit as proposed by \cite[Kruijssen, Dale \& Longmore (2015)]{kruijssen_2015} for gas clouds in the inner $\sim$\,100pc of the CMZ, and the relation of the observed clouds and this model \cite[(Longmore et al. 2013b, Rathborne et al. 2014a; Rathborne et al. 2014b; Rathborne et al. 2015; Kruijssen, Dale \& Longmore 2015; Henshaw et al. 2015; Walker et al. 2015)]{longmore_2013b, henshaw_2015, rathborne_2014a, rathborne_2014b, rathborne_2015, kruijssen_2015, walker_2015}.}
   \label{fig1}
\end{center}
\end{figure}

We measure the infrared luminosities integrated across the CMZ ($|l|$$<$1$^{\circ}$, $|b|<$0.5$^{\circ}$) from {\it Spitzer} 24\,\micron, {\it Herschel} 70\,\micron, and total infrared maps, derive star formation rates (SFRs) and compare to values in the literature \cite[(Yusef-Zadeh et al. 2009, Immer et al. 2012, Longmore et al. 2013, Koepferl et al. 2015)]{yusef_2009, immer_2012, longmore_2013, koepferl_2015}. We find that the global SFR measurements span $\sim$\,$0.05-0.15$\,\solyr. Given that the observational and systematic uncertainties on these measurements are around a factor of two, all of these measurements are deemed to be in good agreement. The implication of this are as follows.

\begin{itemize} 

\item[i)] Using the global properties of the CMZ, SF models over-predict the {\bf \emph{current}} global SFR by an order of magnitude \cite[(Longmore et al. 2013)]{longmore_2013}.

\item[ii)] As the mean time-scales covered by the adopted SFR tracers are $\sim$\,$0.5 - 5$\,Myr \cite[(Kennicutt \& Evans 2012)]{kennicutt_2012}, these results are consistent with recent models predicting that the SFR in the CMZ is episodic on a much longer time-scale of $\sim$\,$10 - 20$\,Myr \cite[(Kruijssen et al. 2014; Krumholz, Kruijssen \& Crocker 2016)]{kruijssen_2014, krumholz_2016}.

\end{itemize} 

\vspace{-6mm}
\section{Small scale: star formation rates on 1\,pc (cloud) scales}\label{small_scale}

We measure the infrared (bolometric) luminosity each gas cloud to determine the embedded stellar population. The SFRs and efficiencies have been estimated assuming SF began in these clouds at pericentre passage with Sgr A* \cite[(Kruijssen, Dale \& Longmore 2015; see Figure 1 and references)]{kruijssen_2015}. These SFRs and efficiencies are compared to several of the commonly used models for SF, assuming the physical properties of the ``brick'' cloud can be used as the initial conditions for SF. We find $\sim$\,$1-4$\,per cent of the cloud mass is converted to stars per free-fall time, which is consistent with ``volumetric'' SF models.

\smallskip

The results shown here are consistent with the idea that the cause of the low SFR in the bulk of the gas is due to it not being gravitationally-bound despite its very high density \cite[(Kruijssen et al. 2014; Krumholz \& Kruijssen 2015)]{kruijssen_2014, krumholz_2015}.

\end{document}